\documentclass[11pt,a4paper]{article}
\usepackage{jheppub}
\usepackage{caption}
\usepackage{subcaption}
\usepackage{comment}

\newcommand{\fig}[1]{Fig.\ \ref{#1}}
\newcommand{\figtwo}[2]{Fig.\ \ref{#1} (\subref{#2})}

\newcommand{\pt}{$p_T$}
\newcommand{\eq}[1]{Eq.~(\ref{#1})}
\newcommand{\infinity}{\infty}

\newcommand{\xtherm}{$x_{therm}$}
\newcommand{\sect}[1]{Section #1}

\title {Strong-coupling Jet Energy Loss from AdS/CFT}
\author{R.\ Morad}
\author{and W.\ A.\ Horowitz}

\affiliation{Department of Physics\\ University of Cape Town \\ Private Bag X3, Rondebosch 7701, South Africa}
\emailAdd{razieh.morad@uct.ac.za}
\emailAdd{wa.horowitz@uct.ac.za}

\abstract{%
We propose a novel definition of a holographic light hadron jet and consider the phenomenological consequences, including the very first fully self-consistent, completely strong-coupling calculation of the jet nuclear modification factor $R_{AA}$, which we find compares surprisingly well with recent preliminary data from LHC.  We show that the thermalization distance for light parton jets is an extremely sensitive function of the \emph{a priori} unspecified string initial conditions and that worldsheets corresponding to non-asymptotic energy jets are not well approximated by a collection of null geodesics.  Our new string jet prescription, which is defined by a separation of scales from plasma to jet, leads to the re-emergence of the late-time Bragg peak in the instantaneous jet energy loss rate; unlike heavy quarks, the energy loss rate is unusually sensitive to the very definition of the string theory object itself.  A straightforward application of the new jet definition leads to significant jet quenching, even in the absence of plasma.  By renormalizing the in-medium suppression by that in the vacuum we find qualitative agreement with preliminary CMS $R_{AA}^{jet}(p_T)$ data in our simple plasma brick model.  We close with comments on our results and an outlook on future work.
}


\begin{document}

\maketitle

\section{Introduction}
The spectacular measurements from the Relativistic Heavy Ion Collider (RHIC) \cite{Adcox:2004mh,Adams:2005dq,Adare:2006nq,Abelev:2006db} and the Large Hadron Collider (LHC) \cite{Aamodt:2010pa,Aamodt:2010jd,ALICE:2012ab,Aad:2010bu,ATLAS:2012at,Aad:2013xma,Chatrchyan:2011sx,Chatrchyan:2011pe,Chatrchyan:2012np,CMS:2012aa,Chatrchyan:2012ta} provide compelling evidence for the man-made creation of quark-gluon plasma (QGP), the deconfined state of QCD matter at temperatures above $\sim160$ MeV.  The challenge to the high-energy nuclear physics community is to describe and understand the non-trivial non-Abelian emergent many-body physics properties of this novel form of matter not seen in the universe since a microsecond after the Big Bang.  The challenge is formidable, as the properties of the QGP appear to be far less simple than originally anticipated \cite{Gyulassy:2004zy}: the medium rapidly thermalizes in $\sim1$ fm \cite{Hirano:2005xf}; appears \cite{Schenke:2010rr} to be nearly perfect \cite{Kovtun:2004de}, with an extremely low viscosity-to-entropy ratio $\eta/s\,\sim\,1/4\pi$; and is surprisingly transparent to high-momentum particles \cite{Horowitz:2011gd}.  Is it possible to simultaneously describe these three observations and all the other related data from the collider experiments within a single conceptual framework?

The most stunning result of the past decade of high-energy nuclear physics research is that the first two major observations are most naturally understood within the framework of the anti--de-Sitter/conformal field theory (AdS/CFT) conjecture---i.e., that the QCD matter at scales a few times $\Lambda_\mathrm{QCD}\,\sim\,160$ MeV is strongly coupled \cite{CasalderreySolana:2011us}.  Of course there are many caveats in the application of the AdS/CFT correspondence to heavy ion phenomenology; ignoring the unproven nature of the correspondence, most important, it is difficult to quantify the corrections due to calculations performed in a theory close to but nevertheless different from QCD.  Nevertheless, the application of leading order results from AdS/CFT yield qualitative agreement with the rapid timescale for thermalization \cite{Chesler:2010bi,Heller:2011ju,Heller:2012je} and the size of the entropy-to-viscosity ratio $\eta/s$ \cite{Kovtun:2004de} as extracted from the comparison of predictions from viscous relativistic hydrodynamics models \cite{Hirano:2005xf,Schenke:2010rr} to the momentum space distribution of low-transverse momentum $p_T\lesssim1$ GeV particles measured at RHIC \cite{Adler:2003kt,Adams:2004bi} and LHC \cite{Aamodt:2010pa,ATLAS:2012at,Chatrchyan:2012ta}.  That even very small collisions systems such as p+A can lead to hydrodynamics-like \cite{Bozek:2013uha} collective behavior \cite{CMSpPb} also suggests strong-coupling dynamics.  Simultaneously, the application of the conjecture to the physics of hard probes, that of the third major observation, has been disappointing: leading order energy loss calculations for both light and heavy quarks predict a significant oversuppression of particles compared to the observations at RHIC and LHC \cite{Horowitz:2011cv,Horowitz:2011wm}.

On the other hand, leading order perturbative QCD (pQCD) results appear to naturally describe simultaneously a suite of high-momentum $p_T\gtrsim10$ GeV particle observables from RHIC to LHC \cite{Majumder:2010qh,Horowitz:2012cf,Djordjevic:2014tka}: the magnitude and azimuthal anisotropy of the suppression of light and heavy quarks and gluons at RHIC and LHC as a function of momentum and centrality.  Jet measurements provide another example of the success of the pQCD paradigm in heavy ion collisions.  What a jet ``is'' is inseparable from its experimental definition, but, generally speaking, it is the observation of a clustering of high-momentum particles. In hadronic collisions of protons (or of protons with anti-protons), the measured spectrum of these jets of particles falls off as a power law \cite{Aad:2010ad,CMS:2011ab,Abelev:2013fn}.  This power law production spectrum is quantitatively described by pQCD and is a direct consequence of the QCD coupling becoming weak for large momentum exchanges \cite{Frixione:1995ms,Frixione:1997np,Nagy:2001fj,Nagy:2003tz}.  Measurements of electroweak bosons at RHIC \cite{Adler:2005ig} and LHC \cite{Chatrchyan:2011ua,Chatrchyan:2012vq,Milov:2012pd}, which interact very little with the QGP medium, provide convincing evidence that pQCD correctly describes the production processes in A+A collisions.  Furthermore, the spectrum of jets in A+A collisions is modified but still generally follows a power law \cite{Aiola:2014cja,CMS:2012rba}.  AdS/CFT does not correctly predict this power law behavior.  Hence the dynamics of the earliest times in heavy ion collisions is given by weak-coupling physics, not that of strong-coupling physics.  pQCD-based energy loss models that incorporate the effects of the QGP medium on the evolution of jets \cite{Renk:2013rla,Zapp:2013zya} agree quantitatively with preliminary data \cite{Aiola:2014cja}.  At the same time, even sophisticated higher order calculations have yet to yield a perturbative explanation of the rapid thermalization \cite{Dusling:2010rm,Attems:2012js} and near perfect fluid nature of the QGP medium \cite{Xu:2004mz}.

One may naturally propose that there actually is no tension between the two pictures: due to asymptotic freedom one might naturally expect that observables related to low-momentum particles are best described by a strongly-coupled theory while those associated with a hard momentum scale $p_T\gg\Lambda_{QCD}$ are best described by weak-coupling pQCD.  The problem with this view, however, is that in the energy loss calculations there are several relevant momentum scales in the problem, and it is far from clear which scale(s) dominate the relevant physics.  In particular, energy loss calculations will always involve an explicit temperature scale, and for the foreseeable future collider energies will restrict $T_{QGP}\sim\mathcal{O}(\Lambda_{QCD})$.  Even worse, all perturbative calculations \cite{Wiedemann:2009sh,Majumder:2010qh} assume the bremsstrahlung radiation is composed of quasiparticle quanta.  The result of the calculation is that the vastly most probably energy of the emitted quanta is $E_{rad}\sim\mu_{Debye}\sim gT$.  Since, phenomenologically, $T\sim\Lambda_{QCD}$, $\mu_{Debye}$ should be a strong-coupling scale at which quasiparticles do not exist.  There are several ideas regarding hybrid strong-weak energy loss calculations (see, e.g., \cite{Casalderrey-Solana:2014bpa} for a good discussion and list of references); however, in this work we will pursue the possibility that the non-perturbative dynamics actually dominate the relevant physical processes in energy loss.  The main result of this paper is that we find an agreement between our simple jet suppression model predictions and recent preliminary jet measurements from the CMS collaboration \cite{CMS:2012rba}, suggesting that the single conceptual framework of a strongly-coupled plasma described by the AdS/CFT correspondence might be capable of characterizing the physics of quark-gluon plasma produced in heavy ion collisions.

Our paper is organized as follows.  In \sect{\ref{subsection:fallingstrings}} we give a brief review of the semi-classical string hologram of light quarks in field theory.  We show in \sect{\ref{subsection:energymomentum}} that the thermalization distance for jets in a strongly-coupled plasma depends sensitively on the initial conditions imposed on the string, and demonstrate that the full numerical solution for the string worldsheet for quark jets of $\sim100$ GeV, relevant for heavy ion phenomenology, are not well approximated by a collection of null geodesics.  Confirming the derivation of the instantaneous energy loss rate correction term of \cite{Ficnar:2012np} and the lack of a Bragg peak in the instantaneous energy loss rate for the original holographic jet definition \cite{Chesler:2008uy} in \sect{\ref{subsection:prescription}}, we show that a Bragg peak reappears in the instantaneous energy loss rate for our new jet definition.  Our qualitative results are unchanged for an expanding plasma in \sect{\ref{subsection:expanding}}.  We compute the nuclear modification factor $R_{AA}^{jet}(p_T)$, renormalize the quantity, and compare the result to the preliminary CMS data in \sect{\ref{section:raa}}.  We close with Conclusions and Discussion in \sect{\ref{section:conclusions}}.

\section{Light Quark Energy Loss in AdS/CFT}
\label{section:lightquarks}

\subsection{Jets in a Static Plasma}
\label{subsection:fallingstrings}
According to the AdS/CFT correspondence \cite{Aharony:1999ti}, the $\mathcal{N}=4$ SYM theory at constant, uniform temperature is dual to a $10d$ black hole geometry with the AdS-Schwarzschild (AdS-Sch) metric,
\begin{equation}
    ds^2 = \frac{L^2}{u^2}
    \left [-f(u) \, dt^2 + d \mathbf x^2 + \frac{du^2}{f(u)} \right ] , \label{AdS-BH metric}
\end{equation}
where $f(u) \equiv 1-(u/u_h)^4$ is the blackening factor and $L$ is the AdS curvature radius. Four dimensional Minkowski coordinates are denoted by $x_\mu$ and the coordinate $u$ is an inverse radial coordinate. Thus the boundary of the AdS-Sch spacetime is at $u = 0$ and the event horizon is located at $u=u_h$. The temperature of the equilibrium SYM plasma relates to the event horizon as $T \equiv \frac{1}{(\pi u_h)}$.

Fundamental representation quarks added to the $\mathcal N\,{=}\,4$ SYM theory are dual to open strings moving in the $10d$ geometry \cite{Witten:1998qj} attached to D7 branes \cite{Karch:2002sh}. These branes fill the whole $4D$ Minkowski space and extend along the radial coordinate from the boundary at $u=0$ down to a maximum coordinate at $u=u_m$.
The bare mass $M$ of the quark is proportional to $1/u_m$ \cite{Herzog:2006gh}, so for massless quarks the D7 brane fills the whole radial direction. Open strings with both endpoints attached to the D7 brane are dual to quark--anti-quark pairs on the field theory side. Open strings attached to space-filling D7 branes can fall unimpeded toward and then through the event horizon.

The dynamics of the string is governed by the classical Nambu-Goto action
\begin{equation}
    S_{\rm NG} = -T_0 \int d^2\sigma \sqrt{-\gamma} \,,\label{Nambu-Goto Action}
\end{equation}
where $T_0 = \sqrt{\lambda}/(2 \pi L^2)$ is the string tension (recall that $\lambda$ is the 't Hooft coupling and $L$ is the curvature radius of the AdS space); the world sheet coordinates are $\sigma^a$, where $\tau \equiv \sigma^0$ is denoted as the timelike world sheet coordinate and $\sigma \equiv \sigma^1$ is the spatial coordinate; 
and $\gamma \equiv \det \gamma_{ab}$, with $\gamma_{ab}$ the induced world sheet metric. The string profile is parameterized by a set of embedding functions $X^\mu(\tau,\sigma)$ for which
\begin{equation}
    \gamma_{ab} \equiv \partial_a X \cdot \partial_b X
\end{equation}
and
\begin{equation}
-\gamma \equiv - \det \gamma_{ab} = (\dot X \cdot X')^2 - \dot X^2 X'^{\,2} \,,
\end{equation}
where $\dot X^\mu \equiv \partial_\tau X^\mu$ and $X'^\mu \equiv \partial_\sigma X^\mu$.
The equations of motion for the embedding functions are obtained in the usual way by extremizing the action subject to certain boundary conditions.  For open strings, the boundary condition is that no momentum flows from the end of the string, which implies that the string endpoints move transversely to the string at the local speed of light. 

The physical setup of interest is one of a back-to-back jet pair created in a quark-gluon plasma.  We therefore consider configurations for which the string is created at a point and expands in space-time such that the two endpoints of the string move away from each other; the total spatial momentum of the string vanishes. With an appropriate choice of coordinates, in the rest frame of the plasma (equivalent to the rest frame for the whole string) one half of the string has a large spatial momentum in the $+x$ direction while the other half of the string has a large spatial momentum in the $-x$ direction; in this case the embedding function of string $X^{\mu}(\tau,\sigma)$ will be a map to $\left( t(\tau,\,\sigma),\,x(\tau,\,\sigma),\,u(\tau,\,\sigma) \right)$.

\begin{figure}
\center
\includegraphics[scale=0.5]{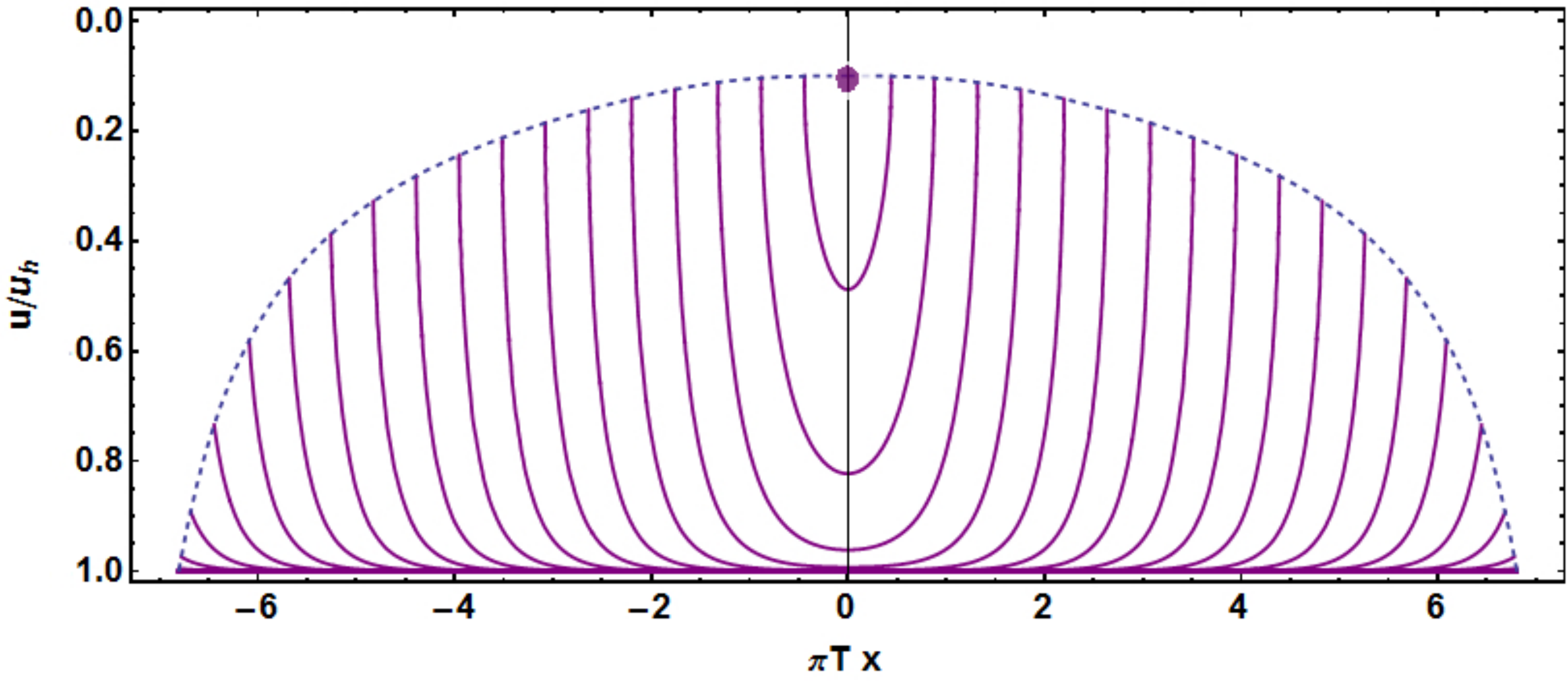}
\caption
  {(Color online) A typical falling string profile obtained numerically. Each purple line shows the string at a different instant in time. The string is created at a point at $u_c=0.1 \, u_h$ and evolves to an extended object. The endpoints of the string move away each other and fall toward the horizon.
  \label{string-adsbh}}
\end{figure}

The profile of an open string that is created at a point in space at time $t=t_c$ is given by
\begin{equation}
    t(0,\sigma)=t_c\, ,\,\,\, x(0,\sigma)=0\, ,\,\,\,u(0,\sigma)=u_c,
\end{equation}
where $\sigma\in[0,\,\pi]$.  After the creation at time $t_c$, the string evolves from a point into an extended object and the string endpoints fall toward the horizon; see \fig{string-adsbh} for a visualization of the string profile at various times after creation.

For precise numerical studies of the string profile, it is more convenient to use the Polyakov action instead of the Nambu-Goto action \cite{Herzog:2006gh,Chesler:2008wd,Chesler:2008uy}. The Polyakov action is better suited for numerical study because the string's equations of motion become singular whenever the determinant of the induced metric goes to zero; it turns out that the induced metric develops a singularity at late times as the string accelerates toward the black brane \cite{Chesler:2008uy}. With the Polyakov action, one introduces additional degrees of freedom into the problem by allowing a nontrivial worldsheet metric $\eta^{ab}$; with these additional degrees of freedom, one can make the equations of motion well-behaved everywhere on the worldsheet \cite{Herzog:2006gh,Chesler:2008wd,Chesler:2008uy}. The Polyakov action for the string has the form
\begin{equation} \label{Polyakov action}
    S_P=-\frac{T_0}{2}\int d^2\sigma \> \sqrt{-\eta} \, \eta^{ab}\,
    \partial_a X^\mu\partial_b X^\nu \, G_{\mu\nu}\, .
\end{equation}
Varying the Polyakov action with respect to $\eta_{ab}$ generates the constraint equation as follows
\begin{equation} \label{metricEquiv}
    \gamma_{ab}=\frac {1}{2}\,\eta_{ab} \, \eta^{cd} \, \gamma_{cd}\, .
\end{equation}
The Nambu-Goto action can be recovered by substituting Eq.~(\ref{metricEquiv}) into the Polyakov action.
Variation of the Polyakov action with respect to the embedding functions $X^\mu$ leads to the equation of motion
\begin{align}
	\label{stringEoM}
	\partial_{a} \big[ \sqrt{-\eta}\,\eta^{ab} \, G_{\mu\nu}\, \partial_{b}X^{\nu}  \big] & = \frac{1}{2} \sqrt{-\eta}\,\eta^{ab}
    \frac{\partial G_{\nu \rho}}{\partial X^{\mu}} \,
    \partial_{a}X^{\nu}\partial_{b}X^{\rho} \nonumber \\[5pt]
    \Longleftrightarrow \qquad \nabla_a\,  \Pi^a_{\mu} & = -\frac{T_0}{2} \,\eta^{ab}
    \frac{\partial G_{\nu \rho}}{\partial X^{\mu}} \,
    \partial_{a}X^{\nu}\partial_{b}X^{\rho},
\end{align}
where $\Pi^a_\mu$ are the canonical momentum densities associated with the string that are obtained from varying the action with respect to the derivatives of the embedding functions,
\begin{equation} \label{Canonical momentum}
    \Pi^a_{\mu}(\tau,\sigma)
    \equiv
    \frac{1}{\sqrt {-\eta\,}} \,\frac{\delta S_{\rm P}}{\delta ( \partial_a X^\mu(\tau,\sigma))} = -T_0\, \eta^{ab}\, \partial_b X^\nu\, G_{\mu  \nu}.
\end{equation}
The open string boundary conditions are
\begin{equation} \label{openBC}
    \Pi^\sigma_{\mu}(\tau,\sigma^*) =0\,,
\end{equation}
where $\sigma^* \, = \, 0$ or $\pi$ is a string endpoint.

In order to optimize the performance of the numerical integrator, we choose a worldsheet metric
of the form \cite{Herzog:2006gh,Chesler:2008wd,Chesler:2008uy}
\begin{equation} \label{Worldsheet Metric}
    \|\eta_{ab}\|= \left(\begin{array}{cc} -\Sigma(x,u) & 0 \\ 0 & \Sigma(x,u)^{-1}
    \end{array}\right),
\end{equation}
where $\Sigma$ is called a stretching function, which can be a function of $x(\tau,\sigma)$ and $u(\tau,\sigma)$. In fact, the choice of worldsheet metric is a choice of gauge. A common choice is conformal gauge with  $\Sigma=1$. We choose $\Sigma$ such that the singularities in the equations of motion are cancelled (for the specific $\Sigma$'s used in this work, see below).

Having derived the equations of motion, we now need to supply physically relevant, self-consistent initial conditions (IC) for the string profile.  
Self-consistency in this case means the IC satisfy the equations of constraint and the boundary conditions.  Using  Eq.~(\ref{Worldsheet Metric}) in Eq.~(\ref{metricEquiv}) yields the constraint equations
\begin{equation}
    \dot X \cdot X' = 0\,,\,\,\,\,\,\,
    \dot X^2 + \Sigma^2 X'^2 = 0\,\label{con2}.
\end{equation}

The $\sigma$ derivatives of $X^\mu$ are initially zero for the string with point-like IC. So, in order to satisfy \eq{con2} we just need to choose IC that satisfy $\dot X^2=0$ and obey the boundary condition Eq.~(\ref{openBC}). The constraint equations are initially satisfied by the following relation in the AdS-Sch metric,
\begin{equation}
    f\,\dot{t}^{\,2}=\dot{x}^2+\frac{\dot{u}^2}{f}\,\label{con3} .
\end{equation}

To proceed it is useful to express the general formula of the canonical momentum densities associated with the string in the AdS-Sch metric. From Eq.~(\ref{Canonical momentum}), we have
\begin{eqnarray} \label{momentum densities}
    \left(\begin{array}{ccc} \Pi_t^{\tau}  \\ \Pi_x^{\tau} \\  \Pi_u^{\tau}
    \end{array}\right) = \frac{\sqrt \lambda}{2 \pi} \left(\begin{array}{ccc} -\frac{f(u)}{\Sigma\,u^2} \, \dot t \\  \frac{1}{\Sigma\,u^2} \, \dot x \\\frac{1}{\Sigma\,f(u) \, u^2} \, \dot u
    \end{array}\right), 
    \;
    \left(\begin{array}{ccc} \Pi_t^{\sigma}  \\ \Pi_x^{\sigma} \\  \Pi_u^{\sigma}
    \end{array}\right) = \frac{\sqrt \lambda}{2 \pi} \left(\begin{array}{ccc} \frac{f(u) \, \Sigma}{u^2} \, t' \\ -\frac{\Sigma}{u^2} \, x' \\ -\frac{\Sigma}{f(u) \, u^2} \,  u'
    \end{array}\right)
    .
\end{eqnarray}
The open string boundary condition Eq.~(\ref{openBC}) requires $X'^\mu(\tau,\sigma^*)=0$ for all $\tau$. In particular, the open string boundary conditions hold at $\tau = 0$, and we require that our IC satisfy
\begin{equation}
	x'(0,\,\sigma^*) = u'(0,\,\sigma^*) = t'(0,\,\sigma^*) = 0.
\end{equation}
Our solution is then guaranteed to satisfy the boundary conditions for all $\tau$ if we set
\begin{equation} \label{finalBC}
       \dot{x}'(0,\sigma^*)=\dot{u}'(0,\sigma^*)=0.
\end{equation}
(Note that the constraint equation at $\tau = 0$, \eq{con3}, automatically yields $\dot{t}'(0,\sigma^*)=0$ when \eq{finalBC} is satisfied.) 

The next step is to find specific IC that satisfy the constraint \eq{con3} and obey the boundary conditions \eq{finalBC}. We seek IC such that the string is long-lived, has most of its energy and momentum concentrated near its endpoints, and produces stable numerical solutions (some IC yield solutions for which numerical noise builds to uncontrolled fluctuations along the string).  Different IC correspond to different states in the dual field theory on the boundary. IC with a complicated dependence on $\sigma$, including exponential terms, have been studied in, e.g., \cite{Chesler:2008wd}. One set of IC that satisfy our criteria are 
\cite{Chesler:2008uy}
\begin{eqnarray}
   &&  \dot{x}(0,\sigma) =A \, u_{\rm c} \cos \sigma\,, \nonumber \\
    && \dot{u}(0,\sigma) = u_{\rm c}\sqrt{f(u_{\rm c})} \, (1-\cos 2\sigma)\,, \\
   && \dot{t}(0,\sigma) = \frac{u_{\rm c}}{\sqrt{f(u_{\rm c})}} \, \sqrt{A^2\,\cos^2 \sigma+(1-\cos 2\sigma)^2}\, ,\label{eq:IC} \nonumber%
 \end{eqnarray}
where $u_c$ and $A$ are free parameters that can be related to the energy and momentum of the dual quark in the field theory (see below).  The string starts as a zero-length point that blasts apart; see \fig{string-adsbh} for a visualization of the evolution of a typical string.  These IC yield a string profile that is symmetric about $x=0$ at all times, because $\dot{x}(0,\sigma)$ is antisymmetric about $\sigma=\pi/2$ while $\dot{u}(0,\sigma)$ is symmetric. 

While the equation of constraint is satisfied by the IC by construction, a nontrivial consistency check of any numerical solution is that the solution satisfies the equation of constraint for all $\tau$.  We performed this explicit check, and our solutions respected the equation of constraint for all $\tau$.

Now we choose a stretching function such that the equations of motion remain well behaved everywhere on the world sheet. We use stretching functions of the form \cite{Chesler:2008wd}
\begin{equation}
\Sigma(x,u)=
\left(\frac{1-u/u_h}{1-u_{\rm c}/u_h}\right)^a
\left(\frac{u_{\rm c}}{u}\right)^b
\label{eq:stretch}
\end{equation}
and solve the equation of motion Eq.~(\ref{stringEoM}) numerically with \emph{Mathematica}'s \texttt{NDSolve} to obtain the embedding functions $X^{\mu}$ as a function of $(\tau,\sigma)$. We choose the values of $a$ and $b$ case by case; $a$ and $b$ are in the range of 1 to 3. The shape of a representative string solution at different times is depicted in \fig{string-adsbh}. As expected, the two endpoints of the string move away from each other as the string extends along the $x$ direction and falls toward the horizon.

\subsection{Energy, Momentum, and Virtuality of the String}\label{subsection:energymomentum}

Since $G^{\mu\nu}$ depends only on $u$, for $\mu$ corresponding to $(t,\,\vec{x})$ we have 
\begin{equation}
\nabla_a\Pi_\mu^a = 0. \label{covariant EQM}
\end{equation}
Hence the corresponding momentum densities $\Pi^a_\mu$ are conserved Noether currents on the worldsheet associated with the invariance of the action under spacetime translations.  The $\Pi_\mu^a$ describe the flow of the $\mu$ component of the spacetime momentum of the string along the $a$ direction on the worldsheet \cite{Lawrence:1993sg}.

The conserved charges associated with these currents are defined by
\begin{equation}
p_\mu^\gamma \equiv \int\limits_\gamma \ast \Pi_\mu \,\label{currents1} ,
\end{equation}
where $\gamma$ represents a curve on the worldsheet and $p_\mu^\gamma$ is the $\mu$ component of the spacetime momentum that flows through this curve.
For a general curve on the worldsheet $\gamma(\lambda)$, Eq~(\ref{currents1}) can be explicitly written as \cite{Carroll:2004st}
\begin{align}
	p_\mu^\gamma  & = \int\limits_\gamma \ast \Pi_\mu = \int\limits_\gamma \epsilon_{ab} \, \Pi^a_\mu \, d\sigma^b = \int\limits_\gamma \sqrt{-\eta} \, \tilde{\epsilon}_{ab} \, \Pi^a_\mu \, d\sigma^b \nonumber \\
	& = \int\limits_{\lambda_i}^{\lambda_f} \sqrt{-\eta} \, \tilde{\epsilon}_{ab} \, \Pi^a_\mu \, \frac{d\gamma^b}{d\lambda} \, d\lambda,
	\label{eq:lineint}
\end{align}
where $\tilde{\epsilon}_{ab}$ is the usual Levi-Civita symbol.

In the static gauge $\tau = t$ one may readily find the four-momentum of the string at a specific time $t$, which corresponds to the usual four-momentum of the quark--anti-quark pair in the field theory \cite{Ficnar:2012np}.  Taking $\tilde{\epsilon}_{\tau\sigma} = +1$,
\begin{equation}\label{momentum}
p_\mu (t)=\int_0^\pi d\sigma \, \sqrt{-\eta}\,\Pi_\mu^\tau(\sigma,t).
\end{equation}
The total energy of the string is thus
\begin{equation}
E_{\rm string} = p^0 = -p_0 = - \int_0^\pi d \sigma \> \sqrt{-\eta} \, \Pi^\tau_{t}(0,\sigma),
\end{equation}
where $\Pi^\tau_{t}$ denotes the conserved canonical energy density given by \eq{momentum densities}. Substituting \eq{momentum densities} and \eq{eq:IC} into the above equation, the energy as a function of the initial condition parameters $u_c$ and $A$ is
\begin{equation}
E_{\rm string} = \frac{\sqrt{\lambda}}{2 \pi} \,
\frac{f(u_{\rm c})}{\Sigma(x_{\rm c} ,u_{\rm c}) \, u_{\rm c}}
\int_0^{\pi} d \sigma \> \sqrt{A^2\,\cos^2 \sigma+(1-\cos 2\sigma)^2}\,.
\end{equation}
Note that here $\Sigma(x_c,\,u_c) = 1$, but that $\Sigma$ is not necessarily 1 in general at the initial production point.  By symmetry the energy of the quark in the quark--anti-quark pair is half of the string energy; hence
\begin{equation}
E_q \equiv \frac{1}{2} \, E_{\rm string} \,.
\end{equation}
Similarly one may obtain the momentum of the quark (and its gluon cloud) in terms of the parameters of the IC, which gives
\begin{equation}
P_q = \frac{\sqrt \lambda}{2\pi}\frac{A}{u_c\,\Sigma(x_{\rm c} ,u_{\rm c})},
\end{equation}
where we capitalize the momentum of the jet in the field theory to distinguish it from the momentum $p_\mu^\gamma$ in the dual theory.

Now that we have the equations of motion and constraint, the boundary conditions, and a set of reasonable, self-consistent initial conditions, we would like to characterize the resulting worldsheet solutions.  A useful measure of the stopping power of the strongly-coupled plasma is the thermalization distance, \xtherm{}, which is defined as the length along the $x$ direction from the point of production of the original point-like string to the point at which the end of the string falls through the black hole horizon.\footnote{As the string is symmetric, it does not matter which endpoint one follows.  Note that we do not actually determine the exact point at which the string endpoint falls through the black hole horizon as in the coordinates we work in the endpoint only actually falls through the black hole horizon as $t\rightarrow\infinity$.  Rather, we follow the string until the endpoint appears to reach its asymptotic distance from its point of origin.}  On the field theory side of the duality, \xtherm{} corresponds to the length of the plasma traversed before the jet becomes completely thermalized (i.e.\ indistinguishable from the plasma).

\begin{figure}[!htbp]
\centering
\begin{subfigure}[b]{.01in}
    \captionsetup{skip=-15pt,margin=-10pt}
    \caption{}
    \label{subfig:xtherm}
\end{subfigure}
\begin{subfigure}[b]{2.56in}
    \centering
     \includegraphics[width=2.56in]{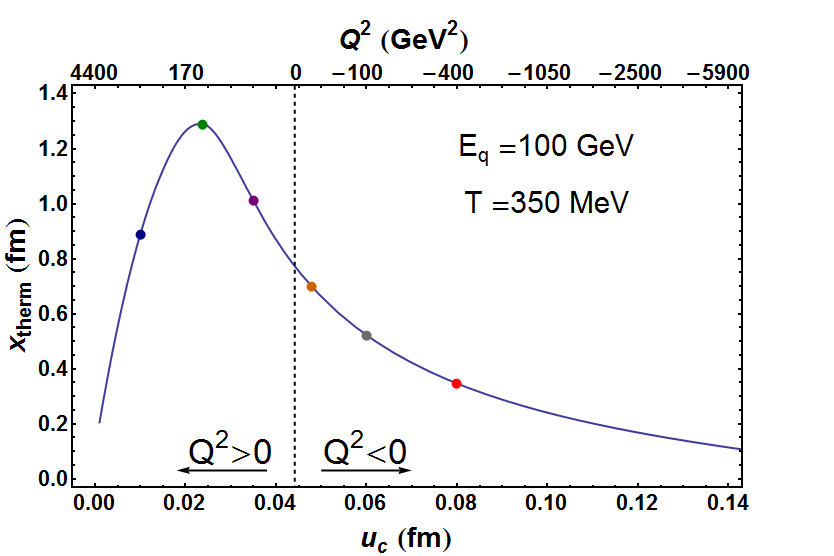}
\end{subfigure}
\begin{subfigure}[b]{.05in}
    \captionsetup{skip=-15pt,margin=-20pt}
    \caption{}
    \label{subfig:endpoint}
\end{subfigure}
\begin{subfigure}[b]{3.214in}
    \centering
    \includegraphics[width=3.214in]{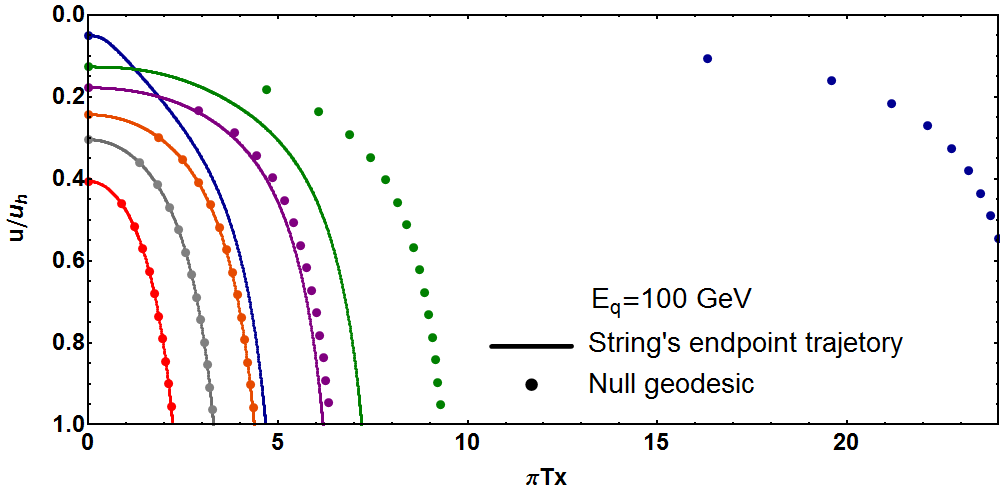}
\end{subfigure}
\vspace{.05in}
\caption{(Color online) (\subref{subfig:xtherm}) The maximum stopping distance \xtherm{} for $E_q = 100$ GeV strings in a $T = 350$ MeV plasma as a function of the creation position of the string in the radial direction $u_c$, which is equivalent to varying the virtuality $Q^2$. The dashed black line corresponds to zero virtuality.  
(\subref{subfig:endpoint}) The trajectory of the string endpoint (solid line) compared with the null geodesic (dots) for different sets of initial conditions for an $E_q = 100$ GeV string. The different IC are represented by dots of the same color in the \xtherm{} plot of (\subref{subfig:xtherm}).}
\label{endpoint-virtuality}
\end{figure}

In \figtwo{endpoint-virtuality}{subfig:xtherm}, we plot a distribution of \xtherm{} for a 100 GeV jet for a variety of values of $u_c$ and $A$; we use $\lambda = 5.5$ \cite{Gubser:2006nz} throughout the paper.  It is useful to translate the IC parameters $u_c$ and $A$ into the virtuality of the jet in the field theory, which we define as
\begin{equation}
	Q^2 \equiv E_q^2 - P_q^2 \,;
\end{equation}
we will use this particle physics sign convention for $Q^2$ throughout the paper.

Using this definition of $Q^2$ we also plot our \xtherm{} distribution against the corresponding $Q^2$ virtuality.\footnote{Note that the first light quark energy loss paper \cite{Chesler:2008uy}, from which we took our string IC, also explored the 2D input parameter space of $u_c$ and $A$ but claimed that the dual quark was always on-shell.}  Notice the huge factor of $\sim\!10$ difference in the thermalization distance depending on the precise choice of parameters used with our IC \eq{eq:IC}.

Recent work \cite{Chesler:2014jva} examined the consequences for jet energy loss in a strongly-coupled plasma by approximating the string in the dual theory as a collection of points; these points then evolved along null geodesics.  We show in \figtwo{endpoint-virtuality}{subfig:endpoint} a comparison between the trajectory of the endpoint of our string and the null geodesic representing the endpoint of the string according to the prescription of \cite{Chesler:2014jva}.  We chose 5 representative values of $Q^2$ for the $E_q = 100$ GeV jet for the comparison; the exact parameters for the IC, equivalently the precise values of $Q^2$, are represented by dots on the \xtherm{} curve in \figtwo{endpoint-virtuality}{subfig:xtherm}.  Unlike at asymptotic energies, at energies accessible with current collider technologies one can see that the validity of the null geodesic approximation to the endpoint trajectory of the string also depends sensitively on the IC of the string.  In particular, at $E = 100$ GeV the approximation is only valid for $Q^2 < 0$.  

\begin{figure}[!htbp]
	\hspace{0.5in}
	\begin{subfigure}[b]{5.06in}
		\includegraphics[width=5.06in]{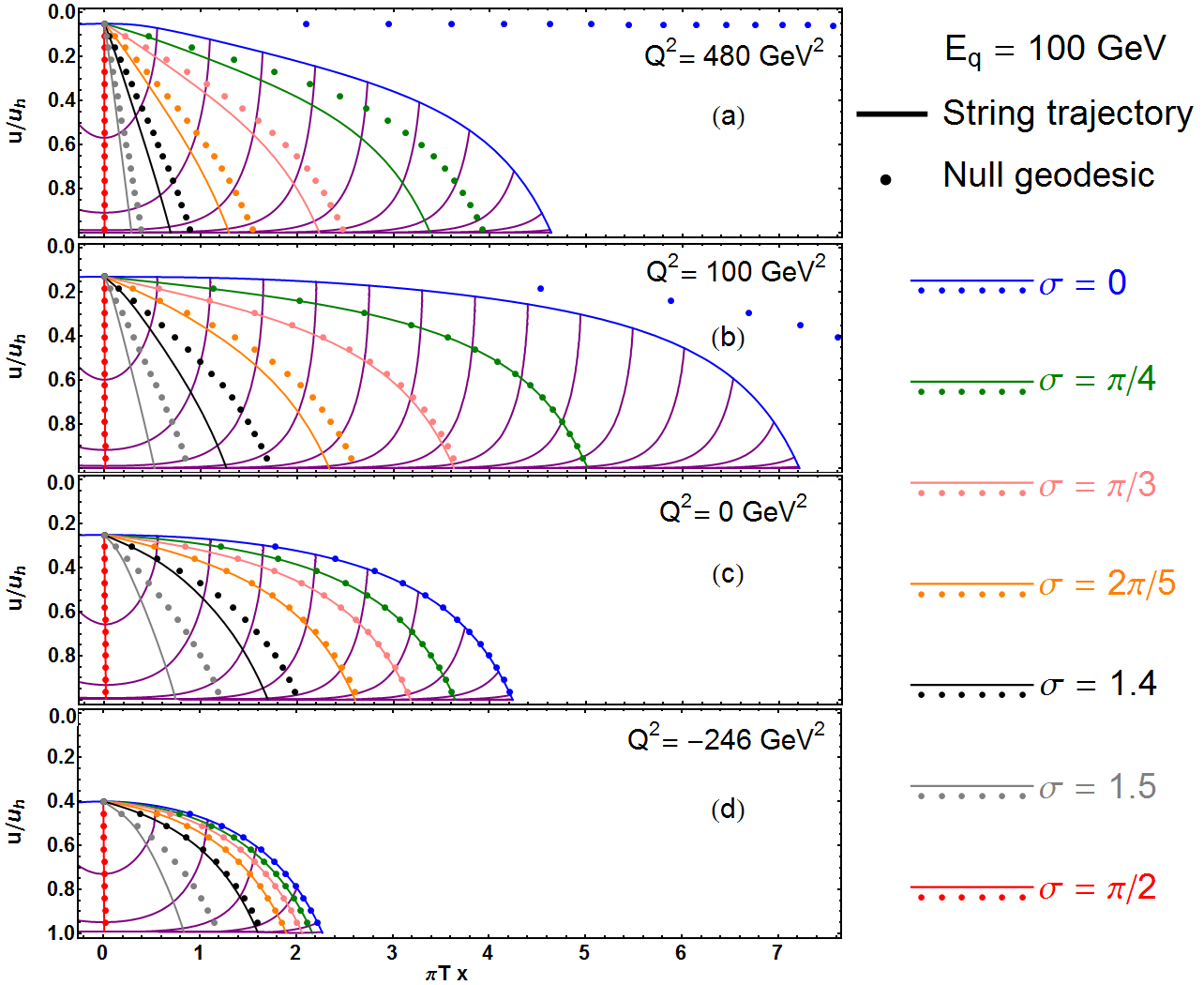}
		\setcounter{subfigure}{1}
		\label{subfig:a}
	\end{subfigure}\\
	\begin{subfigure}[b]{0.425in}
		\addtocounter{subfigure}{3}
		\label{subfig:d}
		$\phantom{A}$
	\end{subfigure}
	\begin{subfigure}[b]{3.333in}
		\addtocounter{subfigure}{1}
		\label{subfig:e}
		\includegraphics[width=3.333in]{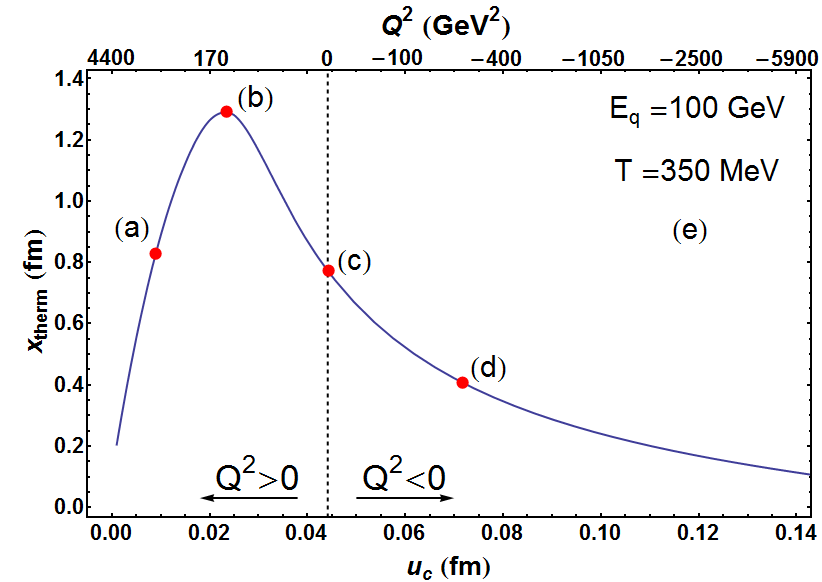}
	\end{subfigure}
	\caption{(Color online) (\subref{subfig:a}) -- (\subref{subfig:d}) The trajectory of different points on the string compared with the null geodesic with the same inclination for a string of $E_q = 100$ GeV and various values of $Q^2$; the values of $Q^2$ are represented by dots on the \xtherm{} plot of (\subref{subfig:e}).}
    \label{fig3}
\end{figure}

In order to further investigate the null geodesic approximation to the full string trajectory, we plot in \fig{fig3} a comparison between the trajectory of different parts of the string with the corresponding null geodesic as per the prescription of \cite{Chesler:2014jva}.  We again used a 100 GeV jet and varied its $Q^2$; the corresponding values of $Q^2$ are represented visually on the \xtherm{} plot included in the figure.  The temperature of the plasma is 350 MeV.  First, notice that for the $Q^2<0$ jet, the good approximation of the $\sigma=0$ trajectory by the null geodesic does not hold for all $\sigma$: as $\sigma$ increases, the approximation becomes worse and is quite poor for $\sigma = 1.5$.  Surprisingly the goodness of the null geodesic approximation can be a complicated function of $\sigma$ and is usually not a monotonic function.  For example, for the $Q^2 = 100$ GeV$^2$ jet the endpoint ($\sigma = 0$) is not well approximated by the null geodesic, the $\sigma = \pi/4$ part of the string is extremely well approximated by a null geodesic, then the approximation gets worse.  Note that the apparent perfect coincidence for the $\sigma = \pi/2$ part of the string with the null geodesic is an artifact of not displaying the temporal dependence; the null goedesic races to the black hole horizon much faster than the portion of string.

\subsection{Jet Prescription and Energy Loss}\label{subsection:prescription}

One may use the thermalization distance of jets to create an extremely crude energy loss model based on \xtherm{} for comparison with the experimentally measured suppression pattern of single inclusive particles fragmented from high-momentum light quarks and gluons \cite{Horowitz:2011cv}.  The naivety of the model yields extremely large theoretical uncertainties; additionally, the string setup much more naturally yields results for jets instead of single particles.  Nevertheless, the theoretical results \cite{Horowitz:2011cv} are consistent with the experimental measurements within the uncertainties.  Encouraged, we wish to have a more theoretically precise prediction of an energy loss observable for comparison to experiment.   

To do so, we need to determine the correct object to investigate on the stringy side of the duality.  This problem is much easier for the heavy quark drag setup \cite{Herzog:2006gh,Gubser:2006bz} if one makes the assumption that the mass of the heavy quark is fundamental, as opposed to generated by the quark's gluon field \cite{Horowitz:2009pw}: those heavy quarks represented by a hanging string in the dual theory unambiguously turn into $D$ and $B$ mesons and their decay products that are ultimately measured by experiments.  The light quark setup is more problematic: there is no clear distinction between the probe and the medium, which is to say that the proper experimental observable to compare to is jets.  
One then has the problem of finding the proper object in the dual string theory that corresponds to a jet, a slippery object even in field theory; jets are truly only defined by the algorithm used to measure them.  Presumably the ideal way to compute jet observables in the dual theory is to compute the energy momentum tensor associated with a high-momentum probe and ``run'' a jet finding algorithm on the result.  Assuming the string worldsheet can be approximated by a collection of null geodesics, the work of \cite{Chesler:2014jva} represents a first attempt at this strategy.  We are currently investigating the possibility of computing the energy momentum tensor from the full numerical string solution, a highly nontrivial work in progress that we hope to report on in a later publication.  

\begin{figure}
\centering
\includegraphics[width=5in]{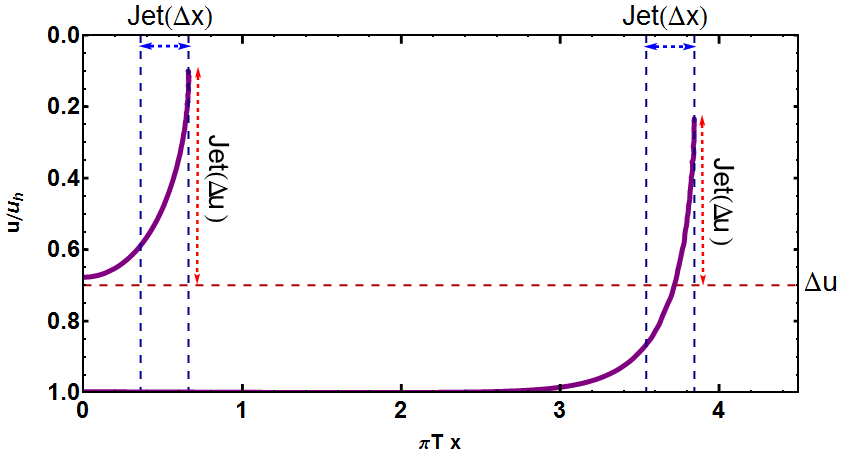}
\caption
{(Color online) Illustration of the $\Delta x$ and $\Delta u$ prescriptions of a jet in the string theory; see text for details. 
  \label{jet} }
\end{figure}

In lieu of the calculation of the energy momentum tensor, previous work \cite{Chesler:2008uy,Ficnar:2012np} relied on using a simpler prescription to approximate the jet results in the dual string theory.  The original suggestion \cite{Chesler:2008uy} defined all of the string within some distance $\Delta x$ of the string endpoint as ``the jet''; see \fig{jet}.  The prescription claims, then, that the energy and momentum of the jet in the field theory is well approximated by the energy and momentum of the part of the string from the string endpoint to the point on the string a distance $\Delta x$ away from the endpoint; the energy and momentum in the string theory is found by integrating the canonical momentum densities \eq{Canonical momentum} from the endpoint to the point on the string a distance $\Delta x$ away in the $x$ direction.  Although the total energy and momentum of each half of the string is independently conserved, the jet is defined as less than half of the string; therefore momentum can flow out of the part of the string encompassed by the definition and into the plasma.  

A major disadvantage of the $\Delta x$ prescription is that it does not connect particularly naturally with any experimental measurement of a jet, which is usually defined by the particles that are measured within some cone in angular and rapidity space.  In particular, even portions of the string that are only infinitesimally above the black hole horizon---and hence are actually indistinguishable from the plasma background---``count'' towards the jet.  In fact, using the $\Delta x$ prescription, a jet that has reached \xtherm{} and is completely thermalized still has a significant, non-zero fraction of its original energy.

Motivated by the separation of energy scales in, e.g., thermal field theory, we propose rather a $\Delta u$ prescription (see \fig{jet}), which we believe will ultimately provide a closer approximation to the result of a more complete calculation.  Since the radial coordinate in the string theory sets an energy scale in the field theory, in our $\Delta u$ prescription the portion of the string above some cutoff $u=u^*$ in the radial direction is considered part of the jet; the portion of the string below the cutoff is considered part of the thermalized medium.  By choosing any value of $u$ above the black hole horizon as the cutoff, we regain the natural result that a jet that is thermalized no longer has detectable energy or momentum.

Given a jet prescription such as either of the above, we may then compute the final momenta of a spectrum of strings in an energy loss model, make predictions for an observable such as the nuclear modification factor $R_{AA}$, and then compare to data.  It is instructive, though, to first examine and compare the instantaneous energy loss rate for the two prescriptions.  

\begin{figure}[!htbp]
	\centering
	\includegraphics[width=2in]{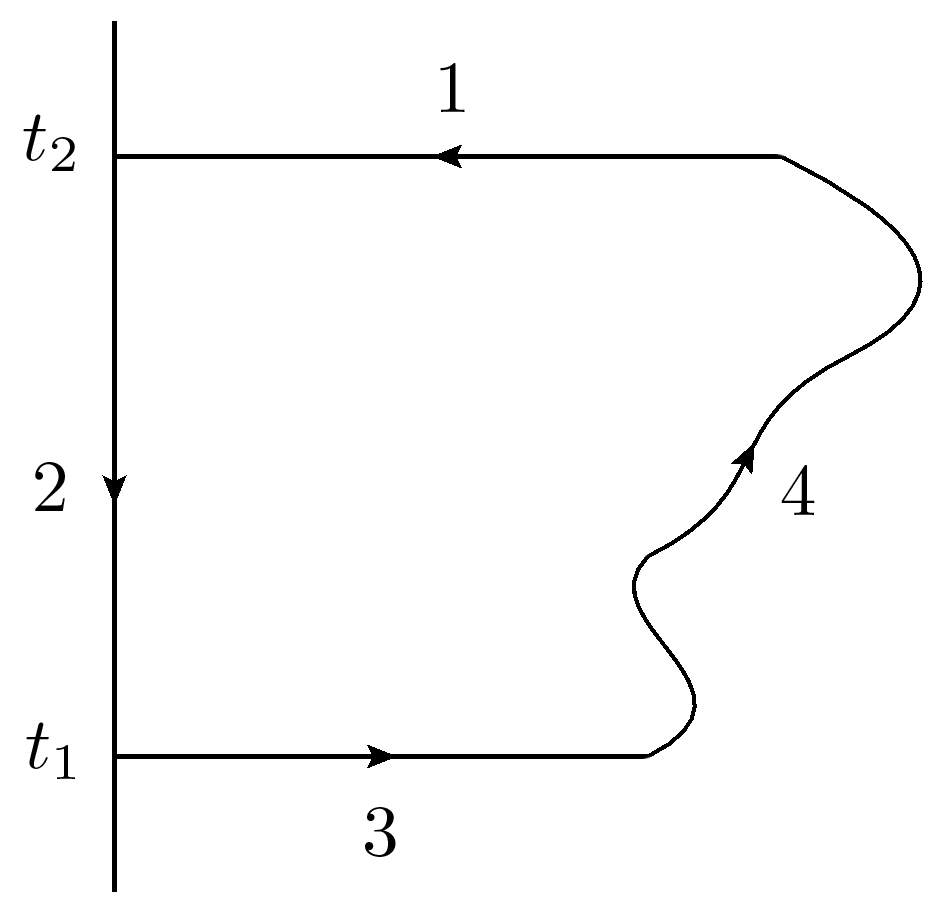}
	\caption{Closed curve $\gamma(\lambda)$ used to derive the instantaneous energy loss for a quark in the dual string theory.  The vertical axis corresponds to the string endpoint $\sigma^* = 0$; time increases going up the axis.  $\sigma$ increases to the right with $\sigma = \pi/2$ and higher not shown.}
	\label{fig:lossderivation}
\end{figure}

Since the momentum densities are conserved \eq{covariant EQM} one may use the relation
\begin{equation}
	\int_\Omega d\,\Pi_\mu = \oint_{\partial\Omega} *\Pi_\mu = 0
\end{equation}
and the curve shown in \fig{fig:lossderivation} to find the instantaneous energy or momentum lost by the jet.  We again work in the static gauge with $\tau = t$ in order to make contact with the momentum of the jet in the field theory at any specific time $t$.  The relevant region of the string worldsheet is bounded by the curves of constant times $t_1$ and $t_2$ of interest, the string endpoint $\sigma^* = 0$, and some curve closing out the region that we take as some general $\gamma_4(t)\equiv\big(t,\,\sigma_\kappa(t)\big)$.  Taking $\gamma_4$ as a general curve is necessary as either of the above jet prescriptions yields a curve $\sigma_\kappa(t)$ on the string worldsheet that is not necessarily a constant in time.  Using the equation for the general line integral of a Hodge dual, \eq{eq:lineint}, we have that
\begin{align}
	0 & = \int_{\gamma_1}*\Pi_\mu + \int_{\gamma_2}*\Pi_\mu + \int_{\gamma_3}*\Pi_\mu + \int_{\gamma_4}*\Pi_\mu \nonumber \\
	0 & = \int_{\sigma_\kappa(t_2)}^0 d\sigma\sqrt{-\eta}\,\Pi_\mu^t + 0 + \int_0^{\sigma_\kappa(t_1)}d\sigma\sqrt{-\eta}\,\Pi_\mu^t + \int_{t_1}^{t_2}dt\sqrt{-\eta}\left(\Pi_\mu^\sigma - \Pi_\mu^t \, \dot{\sigma}_\kappa\right) \nonumber\\
	\Rightarrow & \qquad p_\mu(t_2) - p_\mu(t_1) = -\int_{t_1}^{t_2}dt\sqrt{-\eta}\left(\Pi_\mu^\sigma - \Pi_\mu^t \, \dot{\sigma}_\kappa\right).
\end{align}
To get from the first line to the second line we used the open string boundary condition \eq{openBC} to drop the identically zero contribution from the momentum flow out of the string endpoint at $\sigma^* = 0$ along $\gamma_2$.  In the last line, we used the definition of the momentum \eq{momentum} to rewrite two of the integrals in terms of the quark momentum.  The instantaneous momentum loss is found by taking $t_2 = t_1 + dt$,
\begin{equation}
	\label{eq:instloss}
	\frac{dp_\mu}{dt} =  -\sqrt{-\eta}\left(\Pi_\mu^\sigma - \Pi_\mu^t \, \dot{\sigma}_\kappa\right).
\end{equation}
Our calculation confirms the results of \cite{Ficnar:2012np} and the need for a correction term for the original result \cite{Chesler:2008uy}, perhaps with a more clear derivation.  

The equations of motion and the canonical momenta are naturally functions of $\tau$ and $\sigma$, so it is numerically easier to compute the instantaneous momentum loss in the $(\tau,\,\sigma)$ coordinate system, ultimately evaluating $\tau(t,\,\sigma)$.  The change of coordinates modifies \eq{eq:instloss} to
\begin{equation}
	\frac{dp_t}{dt} = -\frac{\sqrt{-\eta}}{\dot t}\, \left[ \Pi_t^{\sigma} -  \left( \dot t \, \Pi_t^{\tau}+t' \, \Pi_t^{\sigma} \right)\frac{d\sigma_\kappa}{dt} \right]_{\left(\tau\left(t,\sigma_\kappa(t)\right),\,\sigma_\kappa(t)\right)}\, \label{EL-Fic}.	
\end{equation}

\begin{figure}[!ht]
        {\centering
         \includegraphics[width=0.47\textwidth]{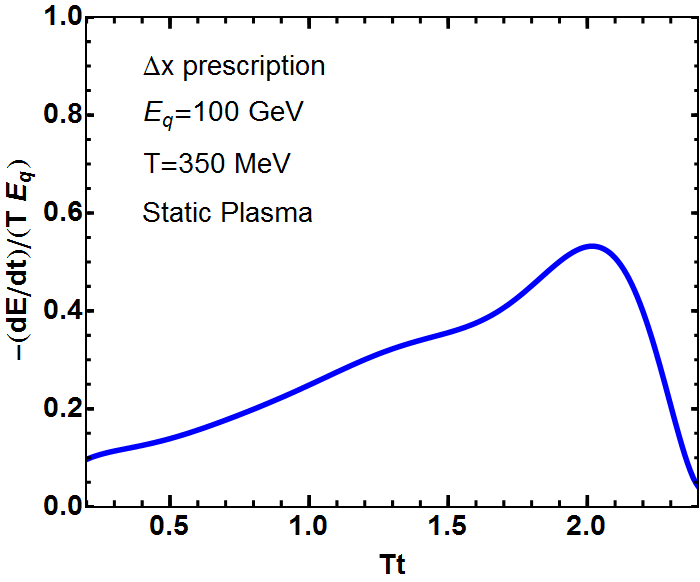}}
    {
  \hspace{0.01\textwidth}
        \includegraphics[width=0.47\textwidth]{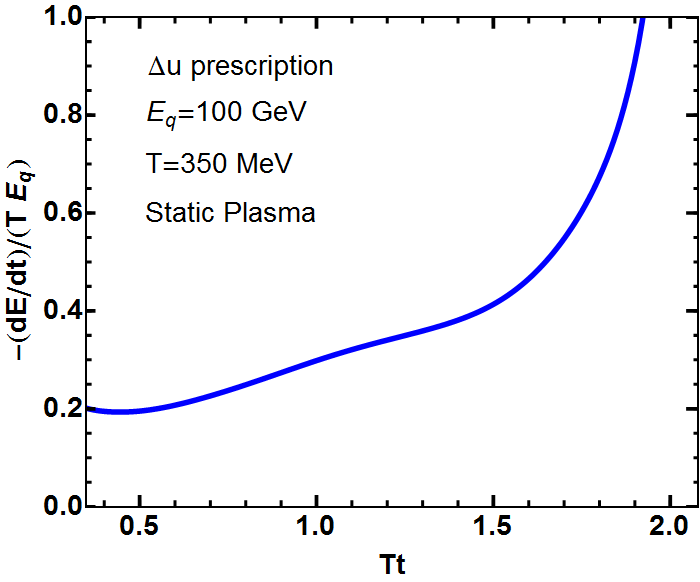}}
            \caption{(Color online) The instantaneous energy loss of a light quark jet as a function of time in the AdS-Sch metric in the $\Delta x$ prescription (left graph) and $\Delta u$ prescription (right graph). 
            The normalization constant $E_q = 100$ GeV is the initial energy of the jet, which has a virtuality of 175 GeV$^2$, and $T = 350$ MeV is the temperature of the plasma. 
            Note the false 0 on the $x$-axis; there is no energy loss for times earlier than those shown.
            }
    \label{EL-AdS}
  \end{figure}

\fig{EL-AdS} plots the instantaneous energy loss curves from implementing \eq{EL-Fic} for both the original $\Delta x$ and our novel $\Delta u$ prescriptions for a 100 GeV jet in a $T = 350$ MeV static plasma; we took $u_c = 0.1u_h$, which gives a $Q^2 = 175$ GeV$^2$ for the jet.  For the $\Delta x$ prescription we took $\Delta x = 0.3/\pi T$ and for the $\Delta u$ prescription $u^*$ corresponding to 500 MeV, $\mathcal{O}(T_{plasma})$, as a reasonable order of magnitude cutoff on the momentum of objects detectable as part of a jet at LHC.  Notice that, consistent with \cite{Ficnar:2012np}, we find that with the correction term the $\Delta x$ prescription of \cite{Chesler:2008uy} yields an instantaneous energy loss that does not have a late-time Bragg peak.  
With our $\Delta u$ prescription the late-time Bragg peak reappears.  It is worth noting that the null geodesic energy-momentum tensor results in \cite{Chesler:2014jva} also show the reappearance of the late-time Bragg peak, which we take as circumstantial evidence supporting our claim that the $\Delta u$ prescription is a reasonable approximation to the full energy-momentum tensor result.

\subsection{Light Quark Energy Loss in an Expanding Plasma}\label{subsection:expanding}

The quark-gluon plasma produced in ultra-relativistic heavy ion collisions is an expanding medium.  Evidence suggests \cite{Hirano:2005xf,Schenke:2010rr} that the dominant growth of the plasma fireball is a one dimensional Hubble expansion along the axis of the beampipe at approximately the speed of light, which is known as Bjorken expansion \cite{Bjorken:1982qr}.  As the plasma expands it adiabatically cools.  The Bjorken expansion gives the dominant contribution to this cooling, with the temperature decreasing like $1/\tau^{1/3}$, where $\tau$ is the proper time in the field theory (defined below).  Since it is likely important in phenomenological studies to capture this time dependence of the temperature of the plasma, we will now investigate the energy loss of light quark jets in a time-dependent dual theory.  Examining the energy loss physics in the time-dependent background has the added benefit that the loss rates will likely be smaller, presumably leading to a better agreement with experimental results.  In this work we use the results of Janik and Peschanski (JP) \cite{Janik:2005zt}.  In the JP metric, the temperature of the plasma in the field theory has (at late times) precisely the time dependence we seek, $T\sim1/\tau^{1/3}$.  

In particular, the JP metric is defined in terms of the proper time and rapidity,
\begin{align}
	x^0 & \equiv \tau \, \cosh y & x^3 & \equiv \tau \, \sinh y \nonumber \\
	\Rightarrow \qquad \tau & = \sqrt{\big(x^0\big)^2 - \big(x^3\big)^2} & y & = \tanh^{-1}(x^3/x^0),
\end{align}
where $x^3$ is defined by the collision beam axis. 
A perfect fluid with energy density $f(\tau)=e_0/\tau^{4/3}$, in the large $\tau$ limit, is dual to the JP metric \cite{Janik:2005zt},
\begin{equation}
\label{JPmetric1}
ds^2=\frac{L^2}{u^2} \left[- \frac{\left( 1-\frac{e_0}{3}
      \frac{u^4}{\tau^{4/3}}\right)^2}{1+\frac{e_0}{3}\frac{u^4}{\tau^{4/3}}} d\tau^2+
\left( {\textstyle 1+\frac{e_0}{3} \frac{u^4}{\tau^{4/3}}}\right) (\tau^2
      dy^2 +dx^2_\perp)+du^2
\right].
\end{equation}
This geometry is similar to the static black hole geometry, but the location of the horizon moves in the bulk as
\begin{equation}
u_h=\left(\frac{3}{e_0}\right)^{1/4}\,\tau^{1/3},
\label{JPhorizon}
\end{equation}
and the temperature of plasma is
\begin{equation}
T(\tau)=\frac{\sqrt 2}{\pi} \, \left(\frac{3}{e_0}\right)^{-1/4}\,\tau^{-1/3}.
\label{T}
\end{equation}

In order to study the light quark energy loss, we use the JP metric in the mid-rapidity limit,
\begin{equation}
\label{JPmetric2}
ds^2=\frac{L^2}{u^2} \left[- \frac{F^2(u,t)}{H(u,t)} dt^2+\,H(u,t)\,dx^2_\perp+du^2
\right],
\end{equation}
where we define $F(u,t)$ and $H(u,t)$ as
\begin{equation}
\label{F-H-definitions}
F(u,t)=1-\left(\frac{u}{u_h}\right)^4 \,, \,\,\,\,\,  H(u,t)=1+\left(\frac{u}{u_h}\right)^4\, .
\end{equation}
As before, the boundary of spacetime is located at $u=0$ and the horizon, $u_h$, moves away from the boundary during the time.

Again, we adopt point-like initial conditions. At the string's creation time, the boundary is at $u=0$, while the horizon is located at $u_h(t_c)$. Note that we can no longer have the $q$-$\bar{q}$ created at $t = 0$ as $u_h(t_c)\rightarrow0$ as $t_c\rightarrow0$; physically, the temperature diverges as $t\rightarrow0$, and the JP approximation breaks down. The constraint equation, \eq{con2}, at the initial time in the JP metric becomes
\begin{equation}\label{constraintJP}
      F^2 \,\dot{t}^{2}=H^2\, \dot{x}^2+H\,\dot{u}^2\, .
\end{equation}
In order to respect the open string boundary conditions, we use the following ansatz for the initial string profile in the JP metric:
\begin{subequations}
\begin{eqnarray}
        x(0,\sigma)&=&0\,\,\,\,\,\,  ,  \,\,\,   \dot{x}(0,\sigma) =A \, u_c \cos \sigma\,, \\
        u(0,\sigma)&=&u_c\,\,\, ,  \,\,\,        \dot{u}(0,\sigma) = u_c\sqrt{H(u_c,t_c)} \, (1-\cos 2\sigma)\,,\\
        t(0,\sigma)&=&t_c\,\,\,\,   ,  \,\,\,\,  \dot{t}(0,\sigma) = u_c \frac{H(u_c,t_c)}{F(u_c,t_c)}\, \sqrt{A^2\,\cos^2 \sigma+(1-\cos 2\sigma)^2}\, .
\end{eqnarray}
\label{IC-JP}%
\end{subequations}
The equations of motion from the Polyakov action in the JP metric can be written as
\begin{subequations}
\begin{eqnarray}
&&\partial_{\tau} \left( \frac{H\, \dot x}{\Sigma \, u^2} \right) -  \partial_{\sigma} \left( \frac{\Sigma \,H \, x'}{ u^2} \right) =0\, ,\\
&&\partial_{\tau} \left( \frac{F^2 \, \dot t}{\Sigma \, H\, u^2} \right) -  \partial_{ \sigma} \left( \frac{\Sigma \, F^2 \, t'}{ H\, u^2} \right) =-\frac{1}{2\, \Sigma} [ \,\, (\dot t^2-\Sigma^2\,t'^2)\,\partial_{t} \left(\frac{-F^2}{H\,u^2}\right) \\ \nonumber
&&~~~~~~~~~~~~~~~~~~~~~~~~~~~~~~~~~~~~~~~~~~~~~~~   +(\dot x^2-\Sigma^2\,x'^2)\,\partial_{t} \left(\frac{H}{\,u^2}\right) ] ,\\
&&\partial_{\tau} \left( \frac{ \dot u}{\Sigma  \, u^2} \right) - \partial_{\sigma} \left( \frac{\Sigma  \, u'}{u^2} \right) =\frac{1}{2\, \Sigma} [ (\dot t^2-\Sigma^2\,t'^2)\,\partial_{u} \left(\frac{-F^2}{H\,u^2}\right)  \\ \nonumber
&&~~~~~~~~~~~~~~~~~~~~~~~~~~~~~~~~~~~~~~~~~~~~~~~   +(\dot x^2-\Sigma^2\,x'^2)\,\partial_{u} \left(\frac{H}{\,u^2}\right) +(\dot u^2-\Sigma^2\,u'^2)\,\partial_{u} \left(\frac{1}{\,u^2}\right) ].
\end{eqnarray}\label{EoMJP}
\end{subequations}

We choose the following stretching function in the JP metric
\begin{equation}
\Sigma(\tau,\sigma)=
\left(\frac{1-u(\tau,\sigma)/u_h(\tau,\sigma)}{1-u_c/u_h(\tau,\sigma)}\right)^a
\left(\frac{u_c}{u(\tau,\sigma)}\right)^b\, \left(\frac{u_h(\tau,\sigma)}{u_h(0,\sigma)}\right)^c\,.
\label{eqstretchJP}
\end{equation}
in order to cancel the singularity of the string metric near the black hole horizon of the JP metric, so the equations of motion remain well-behaved everywhere, especially when parts of the string approach the event horizon. From trial and error we find that the values of $a=3$, and $b,\,c=1.2$  make $\Sigma$ approximately cancel the large factors of $\dot{X}^\mu$ that arise in \eq{EoMJP}, easing numerical evaluation. 

The initial energy of the string in the JP metric is then
\begin{equation}
E_{string}= \frac{\sqrt{\lambda}}{2 \pi} \,
\frac{F(u_c,t_c)}{\Sigma(x_c ,u_c,t_c) \, u_c}
\int_0^{\pi} d \sigma \> \sqrt{A^2\,\cos^2 \sigma+(1-\cos 2\sigma)^2}\,,\label{StringEnergy-JP}
\end{equation}
and the instantaneous energy loss rate for a jet in the JP metric is
\begin{equation}\label{FicnarEL-JP}
\frac{dp_t}{dt}=\frac{\sqrt{\lambda} / 2 \, \pi}{\dot t} \frac{F^2}{H\,u^2}\,\left[ \Sigma \, t'-\frac{d\sigma_\kappa}{dt}  \left(\Sigma \,t'^2-\frac{\dot t^2}{\Sigma} \right)   \right] .
\end{equation}

\begin{figure}[!ht]
        {\centering
         \includegraphics[width=0.47\textwidth]{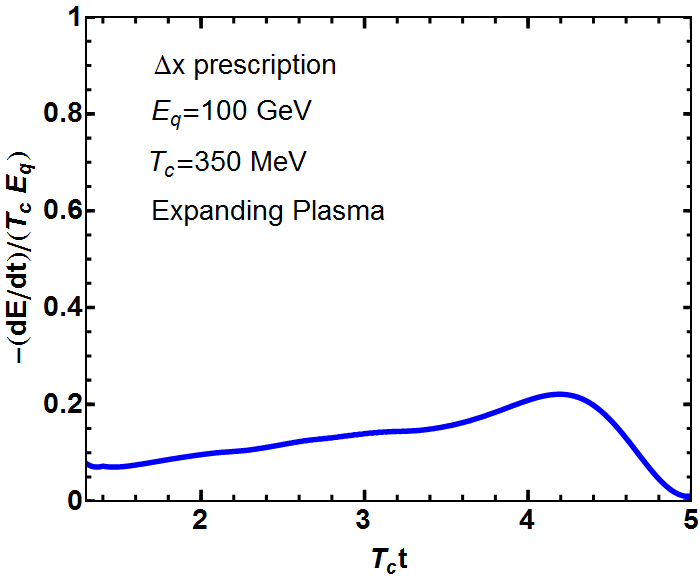}}
    {
     \hspace{0.01\textwidth}
        \includegraphics[width=0.47\textwidth]{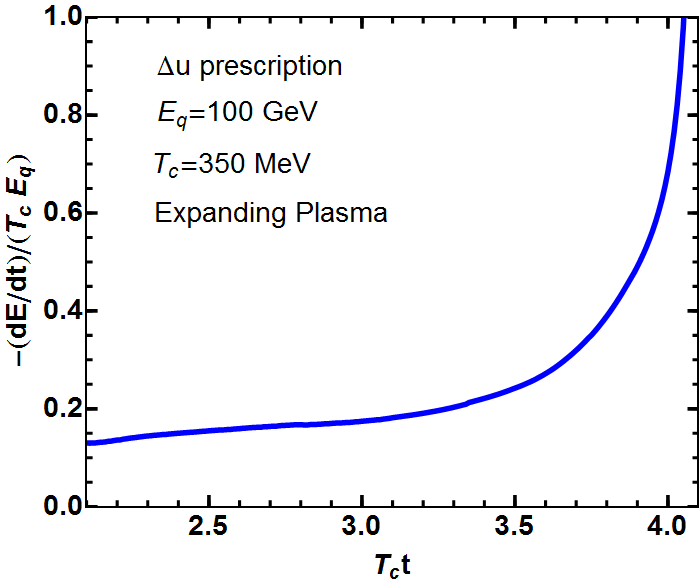}}
            \caption{(Color online) The instantaneous energy loss rate for a light quark jet as a function of time in the JP metric. The left graph is obtained using the $\Delta x$ prescription for a jet while the right graph shows the energy loss from the $\Delta u$ prescription of a jet. The parameters are such that the plasma temperature at the initial time is equal to 350 MeV and the total quark energy is 100 GeV with virtuality of 175 GeV$^2$. Note the false 0 on the $x$-axis; there is no energy loss for times earlier than those shown.}
    \label{EL-JP}
  \end{figure}

We show in \fig{EL-JP} the instantaneous energy loss rates for a 100 GeV jet in a quark-gluon plasma with initial temperature of 350 MeV using the $\Delta x$ and $\Delta u$ prescriptions.  In order to make an apples-to-apples comparison with the AdS-Sch metric results we choose the parameters of the initial profile of string in the JP metric such that the string has the same initial energy and velocity profiles as the string in AdS-Sch metric, whose results we showed in \fig{EL-AdS}. For the $\Delta x$ prescription we set the distance $\Delta x=0.3/\pi\, T_c$ based on $T_c \equiv T(t_c) = 350$ MeV, the initial temperature of plasma in JP metric, which we take the same as the temperature of the static plasma in the AdS-Sch metric. For the $\Delta u$ prescription, we again set our energy scale separating hard and soft physics at 500 MeV. As seen in \fig{EL-JP} the qualitative behavior of both the $\Delta x$ and $\Delta u$ light quark energy loss in the JP metric is the same as the AdS-Sch metric, but the distance the quark travels before thermalizing increases by approximately a factor of 2.

Although we did not explicitly compare full numerical results to the null geodesic approximation in the JP metric, we have no reason to think that there would be a qualitative change in behavior.

\section{Jet Nuclear Modification Factor}\label{section:raa}

Now having a qualitative feel for the thermalization distance and instantaneous energy loss rate from the previous sections, we wish to compare our results to an experimental measurement.  The most natural observable to compare to is the jet nuclear modification factor, $R_{AA}$, which is defined as 
\begin{equation}\label{Defin-RAA}
R_{AA}^{jet}=\frac{dN_{AA \rightarrow jet}(p_T)/dp_T}{N_{bin}dN_{pp \rightarrow jet}(p_T)/dp_T} .
\end{equation}

$R_{AA}$ gives a measure of the effect of the QGP medium on the measurement at hand, in this case jets.  One expects that the number of $pp$-like hard scattering events, those that produce high-\pt{} particles, in a heavy ion collision scales with the number of binary collisions, $N_{bin}$.  Thus if the QGP medium has no effect on the particle(s) involved in a measurement, and assuming the initial configuration of nuclei is approximately that of an incoherent sum of nucleons, then $R_{AA} \simeq 1$.  Hard electroweak probes, predominantly produced in the initial collisions of the nuclei and only weakly interacting with the colored QGP, have $R_{AA}(p_T)\simeq 1$ \cite{Adler:2005ig,Chatrchyan:2011ua,Chatrchyan:2012vq,Milov:2012pd}, thus experimentally confirming the $N_{bin}$ scaling of hard $pp$-like scattering events at RHIC and LHC.  

For a single parton type $R$, which can stand for either a quark $q$ or a gluon $g$, the nuclear modification factor is
\begin{equation}\label{Defin-RAAq}
R_{AA}^{R \rightarrow jet}=\frac{dN_{AA}^{R \rightarrow jet}(p_T)/dp_T}{N_{bin}dN_{pp}^{R \rightarrow jet}(p_T)/dp_T} .
\end{equation}
The experimentally measured jets come from both quarks and gluons, so we must add their contributions together appropriately:
\begin{equation}\label{Defin-RAA2}
R_{AA}^{jet}=\sum_R \,R_{AA}^{R \rightarrow jet} \frac{dN_{pp}^{R \rightarrow jet}(p_T)/dp_T}{ dN_{pp}^{q \rightarrow jet}(p_T)/dp_T+dN_{pp}^{g \rightarrow jet}(p_T)/dp_T} ,
\end{equation}
where we sum the contributions of quarks and gluons jet, $R=(q,g)$.

One may find a relatively simple equation that approximates the partonic $R^{R\rightarrow jet}_{AA}$.  First, take the produced parton to have initial energy $p_T^i$ (we assume the parton is produced at mid-rapidity and only moves in the transverse plane; we also assume that the parton is approximately massless).  The parton then loses a fraction of its energy $\epsilon$ with probability $P\left(\epsilon \, | \, p_T^i,\,L,\,T \right)$, where $L$ is the length of the medium the parton travels through; the parton's final energy is $p_T^f = (1-\epsilon) \, p_T^i$.  The partonic $R_{AA}$ is then \cite{Horowitz:2010dm}
\begin{equation}\label{dNAA}
\frac{dN_{AA}^{R \rightarrow jet}}{dp_T}(p_T^f)=\left\langle \int_0^1 \frac{d \epsilon}{1-\epsilon}  \, \frac{dN_{pp}^{R \rightarrow jet}}{dp_T^R}\Big(\frac{p_T^f}{1-\epsilon}\Big)\,P\Big(\epsilon|\frac{p_T^f}{1-\epsilon},L,T \Big)\right \rangle  ,
\end{equation}
where the angular brackets refer to a geometrical average over the initial production points and angles of emission for the hard partons. %

If one assumes that the AdS energy loss is approximately independent of the initial energy \cite{Horowitz:2010dm} and one only computes the mean energy loss, as we have done in this paper, then
\begin{equation}\label{AdSepsilon}
P\Big(\epsilon \, \Big| \, \frac{p_T^f}{1-\epsilon}, \, L, \, T \Big) \simeq \delta \Big(\epsilon - \epsilon_{AdS}^R \big(p_T^f, \, L, \, T \big)\Big).
\end{equation}
We assume gluons lose their energy by a simple Casimir scaling of the quark energy loss \cite{Horowitz:2010dm},
\begin{equation}\label{AdSepsilong}
\epsilon_{AdS}^g \left(p_T,L,T \right)=\frac{2\,N_c^2}{N_c^2-1}\,\epsilon_{AdS}^q \left(p_T,L,T \right).
\end{equation}

The production spectrum can be well approximated by a power law \cite{Horowitz:2010dm},
\begin{equation}\label{powerlaw}
\frac{dN_{prod}^R(p_T)}{dp_T}=\frac{A}{p_T^{n_R(p_T)}},
\end{equation}
where $A$ is some normalization constant. Assuming a slowly varying power law $n(p_T)$ with respect to $p_T$, we may find a simple equation for the jet nuclear modification factor as follows,
\begin{equation}\label{RAA-AdS}
R_{AA}^{R \rightarrow  jet}(p_T)=\left\langle \int d \epsilon \, P(\epsilon \, | \, p_T, \, L, \, T )  \, \left (1-\epsilon^R \right)^{n_R(p_T)-1}  \right \rangle ,
\end{equation}
where the angular brackets again denote a geometric average.

For a uniform 1D nucleus, the geometric average is an integral over a line of production points with a parton that propagates through the line. In this case, $R_{AA}^{R \rightarrow  jet}(p_T)$ is \cite{Horowitz:2010dm}
\begin{equation}\label{RAA-AdS-simplified}
R_{AA}^{R \rightarrow  jet}(p_T)= \int_0^{L_{max}} \frac{dl}{L_{max}} \, \left (1-\epsilon^R(p_T,\,l,\,T ) \right)^{n_R(p_T)-1}.
\end{equation}

\begin{figure}
	\centering
	\includegraphics[width=4in]{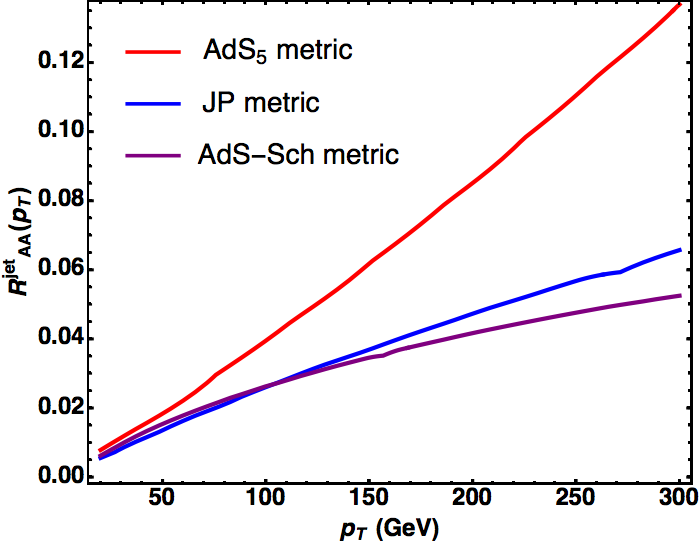}
	\caption
	{(Color online) Jet $R_{AA}$ as a function of $p_T$ for a simple suppression model in the most central Pb-Pb collisions obtained via AdS/CFT strong-coupling energy loss in three different backgrounds. Red, blue, and purple curves show $R_{AA}^{jet}$ for the falling string in the empty $AdS_5$, time-dependent JP and static AdS-Sch metrics, respectively. 
	\label{RAA}}
\end{figure}

In \fig{RAA} we plot $R^{jet}_{AA}$ in a strongly-coupled plasma by using the $\Delta u$ jet energy loss prescription in the AdS-Sch and the JP metrics. The static plasma has a temperature of 350 MeV, and the time-dependent plasma has an initial temperature of 350 MeV at $t_c = 0.6$ fm.  Leading order pQCD gives the production spectrum here for the initial hard quarks and gluons at LHC, $\surd s = 2.76$ TeV \cite{Horowitz:2011gd}.  
We use the most simple toy model for the geometry of the nucleus, taking it to be a 1D object of uniform density of total length $L_{max} = 14$ fm.  As can be seen in \fig{RAA}, the AdS/CFT $R_{AA}(p_T)$ prediction for central collisions at LHC from this very simple model---both from the static plasma AdS-Sch and from the time-dependent JP metric---are significantly oversuppressed compared to the recent preliminary CMS data, which show $R^{jet}_{AA}\sim0.5$ \cite{CMS:2012rba}.

The point-like initial condition falling string that we consider here is dual to the creation of a quark-antiquark pair that flies apart in the strongly coupled plasma, interacting with and losing energy to the plasma. By definition, jets produced in $pp$ collisions do not lose any energy; they propagate in vacuum. Despite this required expectation, one can see from \fig{RAA} that, in using our $\Delta u$ prescription, our jets lose a significant fraction of their energy as they are produced in and propagate through a vacuum ``plasma'' of the same size as that used in the AdS-Sch and JP metrics.  (We find the $R_{AA}^{vacuum}$ by copmuting the string worldsheet in the empty AdS$_5$ metric and keeping $u^*$ at the same numerical value as in the AdS-Sch case.)  

\begin{figure}
	\centering
	\includegraphics[width=4in]{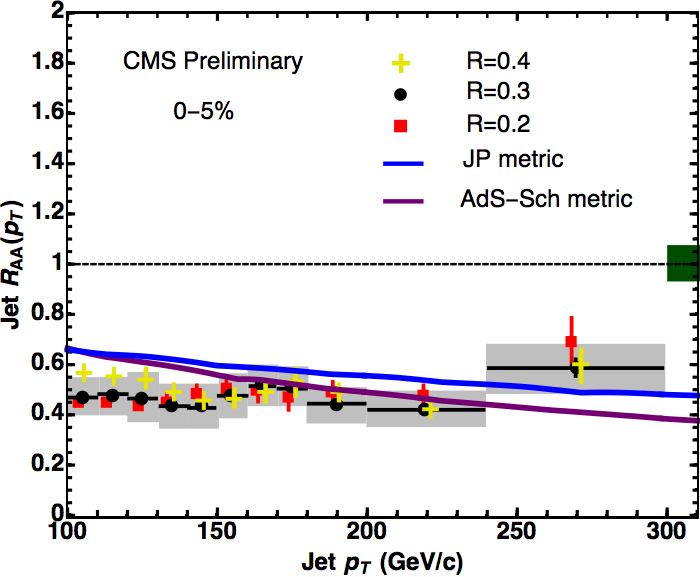}
	\caption{(Color online) AdS/CFT $R_{AA}^{jet}$ as a function of $p_T$ compared with preliminary $\surd s = 2.76$ TeV CMS data \cite{CMS:2012rba} for $0-5$\% central data from LHC.  
	The results of our novel $\Delta u$ prescription calculations in the AdS-Sch and JP metrics are shown by the purple and blue curves, respectively. 
	\label{reg-RAA}}
\end{figure}

Since the experimental $R_{AA}$ measurement is designed to capture the difference between vacuum and plasma physics, we must modify our approach to account for the spurious, large vacuum ``energy loss'' effects stemming from our $\Delta u$ prescription.  We propose that the $\Delta u$ prescription result that should be qualitatively compared with the experimental data is a ``renormalized'' $R_{AA}^{jet}$, which we define as
\begin{equation}\label{eq:renormRAA}
R_{AA}^{renorm} \equiv \frac{R_{AA}^{medium}}{R_{AA}^{vacuum}}.
\end{equation}

We plot the renormalized $R_{AA}^{renorm}$ for jets in both the AdS-Sch and JP metrics in \fig{reg-RAA} and compare with the preliminary CMS data for the most central Pb-Pb collisions at $\sqrt{s_{NN}}=2.76$GeV \cite{CMS:2012rba}. For such a simple energy loss calculation, our results are in surprisingly good agreement with the preliminary CMS measurement.

\section{Discussion and Conclusions}\label{section:conclusions}

In this paper we explored many aspects of jet evolution in strongly-coupled plasma and its phenomenological consequences.  We found that the thermalization distance, the length of plasma through which a jet propagates before fully thermalizing with the medium, is extremely sensitive to the precise initial conditions set for the string; see \figtwo{endpoint-virtuality}{subfig:xtherm}.  Jets in general thermalize very quickly in a strongly-coupled plasma, with extremely short thermalization distances for jets with negative (in the particle physicists' sign convention) or very large positive virtuality.  The thermalization distance is maximized for jets with small positive virtuality.  Perturbative intuition, which must be applicable for the production mechanics---and also likely for some time after---of high-\pt{} jets in particle colliders, suggests that a jet's virtuality is positive and shed in a distance $\sim 1 / Q$.  Hence the string initial conditions relevant for studies related to observables in heavy ion collisions are those of jets with a small, positive virtuality.  However, it is far from clear what a reasonable subset of the multi-infinite dimensional space of initial conditions is to represent the dual to the phenomenologically relevant production of jets in heavy ion collisions; we only explored one dimension of the parameter space for a very specific functional choice for the string initial conditions.  It is necessary, therefore, to find guidance, likely from the weakly-coupled field theory, to narrow down the choices for initial conditions.  We will return to the issue of winnowing down the possible initial conditions in a moment.  

Since there is no yet known string dual to a jet associated with a light parton in a field theory, one must resort to some kind of prescription if one wants to make a comparison to experimental measurements.  The original such prescription defined a jet in the dual theory as all the string within a distance $\Delta x$ of the endpoint of the string \cite{Chesler:2008uy}.  We rather proposed a scale separation between the thermal plasma and the high-\pt{} jet, which we called the $\Delta u$ prescription for short.  In order to further learn about the qualitative physics of our jet definition, we rederived the equations that govern the instantaneous momentum lost along some portion of the string that includes its endpoint, confirming the results of \cite{Ficnar:2012np}.  We also saw no Bragg peak in the energy lost by a $\Delta x$ definition jet \cite{Ficnar:2012np} but found that the Bragg peak reappears when using the $\Delta u$ definition for a jet.  These qualitative insights are true for both a static plasma, \fig{EL-AdS}, and for one that has a time dependence similar to that expected in heavy ion collisions, \fig{EL-JP}.  

Using our novel energy scale separation jet definition, we computed the first fully strongly-coupled nuclear modification factor, $R^{jet}_{AA}(p_T)$, albeit in a highly simplified geometrical model for an ultra-relativistic nucleus-nucleus collision.  We found our simple energy loss model's $R_{AA}^{jet}$ to be highly suppressed, $\lesssim0.1$, in \fig{RAA}, both when using the static plasma AdS-Sch and time-dependent plasma JP metrics.  When we computed the ``$R_{AA}$'' using the vacuum AdS metric we discovered a significant suppression.  Since jets cannot lose energy to a plasma that isn't there, we defined a renormalized $R_{AA}^{jet}$ that we hope correctly captures the relevant dynamical differences in the theory between propagation in vacuum and in medium that lead to the experimental measurements.  \fig{reg-RAA} shows that our renormalized $R_{AA}^{jet}(p_T)$ is in very good agreement with the preliminary CMS measurement of $0-5\%$ central $R_{AA}^{jet}(p_T)$ at LHC \cite{CMS:2012rba}.

Of course one immediately wonders how much confidence to assign to this comparison between the renormalized $R_{AA}^{jet}$  and the experimental measurement and, then, how to proceed.  We checked the robustness of our $R_{AA}^{jets}$ results in two ways.  First, we studied an alternative, subtractive renormalization scheme in which we took 
\begin{equation}
	\Delta E_{AA}^{sub\;ren}(p_T^i,\,L,\,T) \equiv \Delta E_{AA}^{medium}(p_T^i,\,L,\,T) - \Delta E_{AA}^{vacuum}(p_T^i,\,L,\,T).
\end{equation}
Note that in our sign conventions, $\Delta E < 0$.  We found the $R_{AA}^{sub\; ren}$ results qualitatively the same as those found from $R_{AA}^{renorm}$, \eq{eq:renormRAA}.  We also examined the effect on $R_{AA}$ of changing the value of the scale that defines the separation between the hard and soft scales.  Not surprisingly (from the AdS side), the unrenormalized $R_{AA}$'s decreased (increased) with increasing (decreasing) $u^*$.  However, $R_{AA}^{vacuum}$ changed more than $R_{AA}^{medium}$ for any change in $u^*$.  As a result, increasing (decreasing) $u^*$ increased (decreased) $R_{AA}^{renorm}$; i.e., in our strong-coupling approach jets defined by a larger higher momentum particles are less suppressed than jets whose constituents are more medium-like.  It is thus through the renormalization procedure and, hence implicitly due to the string initial conditions, that we recover the expected result on the field theory side of the duality.  

While the agreement shown in \fig{reg-RAA} is at the quantitative level, realistically the comparison is qualitative at best.  Neglecting the obvious differences between QCD and $\mathcal{N} = 4$ SYM, whose effects on the predicted observables are difficult if not impossible to quantify, the nuclear geometry used in the energy loss model is highly oversimplified.  Additionally, as indicated previously, the initial conditions that yield a string solution that is (roughly) equivalent to the jet in an actual collider experiment that enters the plasma at the thermalization time are not known.  In particular, it is not at all clear whether the complete lack of early time energy loss in any of the strongly-coupled jet definitions approximates well the early time jet evolution physics prior to the thermalization of the plasma.  

One glaring omission from our $R_{AA}^{jet}$ discussion is the region of applicability of our calculations and, especially, our renormalization procedure.  One can see from the unrenormalized $R_{AA}^{jet}$ plot in \fig{RAA} that as the jet energy decreases, so do both $R_{AA}^{medium}$ and $R_{AA}^{vacuum}$.  At some point the fraction of vacuum jets that are completely thermalized, an unnatural artifact of the current AdS setup, because so large that it no longer makes sense to multiplicatively renormalize as we have done here (similarly, the subtractive renormalization procedure becomes ill-defined when $\Delta E_{AA}^{medium} \, = \, E$ and, especially, when $\Delta E_{AA}^{vacuum} \, = \, E$).  As can be seen in \fig{RAA}, the fraction of completely thermalized jets increases monotonically as $p_T$ decreases, and there is no natural $p_T$ scale at which to stop trusting our renormalization procedure.  We therefore made the somewhat arbitrary choice to compare only to the higher-$p_T$ preliminary CMS results \cite{CMS:2012rba}, not extending our calculation down to the $p_T$ scales explored by the recent ALICE jet suppression measurement \cite{Reed:2013rpa}.  However, the comparison to the CMS results is sufficient for our purposes here: at the level of our crude energy loss model we qualitatively describe the suppression of $R_{AA}^{jet}$.  In particular, given the robustness of our results with respect to changing renormalization schemes and scale separation values, we are confident that fully strong-coupling dynamics can be used to describe the suppression of high-$p_T$ probes in heavy ion collisions. Further progress in describing experimental results will require significant advances in the understanding of string initial conditions.

That the results of our simple model are in such good agreement with data suggests that we attempt to better define the jet in AdS/CFT and constrain the possible string initial conditions.  We can likely accomplish both goals by computing the energy-momentum tensor associated with the propagation of the classical string solution.  With the energy-momentum tensor in hand, we should be able to compute directly from the string theory the actual quantities measured experimentally.  Strongly-coupled jet production was investigated in \cite{Hatta:2008tx,Arnold:2010ir,Arnold:2011qi}; however, it is clear on theoretical grounds and from experimental measurement that high-momentum particle production in heavy-ion collisions is a weak-coupling process.  One expects perturbative considerations to hold for some non-zero length of time after nuclear overlap, perhaps approximately so even up to the thermalization time of $\tau\sim1$ fm (pQCD-based energy loss calculations \cite{Majumder:2010qh,Horowitz:2012cf,Djordjevic:2014tka} currently assume vacuum evolution of the hard parton before it begins interacting with the medium). One could constrain the string initial conditions by requiring that the resultant energy-momentum tensor at finite time, such as $\tau\,=\,1$ fm, from AdS/CFT give similar results to that from pQCD.  One would then have a hybrid early, weak-coupling/late, strong-coupling physics model for jet quenching in heavy ion collisions.  Under the assumptions in \cite{Chesler:2014jva}, the jet energy-momentum tensor in a strongly-coupled calculation can be relatively easily found by a superposition of contributions from a collection of point particles whose paths approximate the evolution of the string worldsheet.  Unfortunately, we found that at jet energies accessible at current colliders, a collection of null geodesics does not approximate the dynamics of a string worldsheet well; see \figtwo{endpoint-virtuality}{subfig:endpoint} and \fig{fig3}.  
It appears that we are thus left to numerically solve the linearized Einstein's equations with a numerical string as the source, a seemingly highly nontrivial task.

A different unresolved issue is the influence of fluctuations on light probe evolution in a strongly-coupled plasma.  It was shown in \cite{Gubser:2006nz,Horowitz:2014} that these fluctuations play an important role in the implementation of energy loss for heavy quarks in strongly-coupled plasma; determining their role in jet physics is an interesting and important open question.  

The fascinating challenge of pursuing this research is left to future work.

\acknowledgments
The authors wish to thank the South African National Research Foundation, the University of Cape Town, and SA-CERN for support and CERN for hospitality.  The authors also wish to thank Paul Chesler, Andrej Ficnar, Miklos Gyulassy, Krishna Rajagopal, Hesam Soltanpanahi, and Urs Wiedemann for useful discussions.


\providecommand{\href}[2]{#2}\begingroup\raggedright\endgroup

\end{document}